\documentclass[12pt]{article}
\usepackage[margin=2cm]{geometry}
\usepackage{comment}
\usepackage{amsmath,amssymb,extarrows,graphicx,subfigure,setspace}
\usepackage{cite}
\usepackage{slashed}
\usepackage{color}
\makeatother

\newcommand{\be}{\begin{equation}}
\newcommand{\bea}{\begin{eqnarray}}
\newcommand{\eea}{\end{eqnarray}}
\newcommand{\ba}{\begin{array}}
\newcommand{\ea}{\end{array}}
\newcommand{\ee}{\end{equation}}
\newcommand{\bes}{\begin{equation*}}
\newcommand{\beas}{\begin{eqnarray*}}
\newcommand{\eeas}{\end{eqnarray*}}
\newcommand{\bas}{\begin{array*}}
\newcommand{\eas}{\end{array*}}
\newcommand{\ees}{\end{equation*}}

\numberwithin{equation}{section}
\begin{document}
	\onehalfspacing
	\noindent
	
\begin{titlepage}
\vspace{10mm}
\begin{flushright}

\end{flushright}

\vspace*{20mm}
\begin{center}

{\Large {\bf  Complexity and   Behind the  Horizon  Cut Off}\\
}

\vspace*{15mm}
\vspace*{1mm}
Amin Akhavan$^{a,}$, {Mohsen Alishahiha${}^b$, Ali Naseh$^{a}$ and Hamed Zolfi$^{c}$ }

 \vspace*{1cm}

{\it 
 ${}^a$ School of Particles and Accelerators, ${}^b$School of Physics,\\
Institute for Research in Fundamental Sciences (IPM)\\
P.O. Box 19395-5531, Tehran, Iran
\\  
${}^c$ Department of Physics, Sharif University of Technology,\\
P.O. Box 11365-9161, Tehran, Iran
}

 \vspace*{0.5cm}
{E-mails: {\tt amin$_{-}$akhavan,alishah,naseh@ipm.ir,\ hamed.Zolfi@physics.sharif.edu}}%

\vspace*{1cm}
\end{center}

\begin{abstract}
Motivated by $T{\overline T}$ deformation of a conformal field theory we compute holographic 
complexity for a black brane solution  with a cut off using ``complexity=action'' proposal. 
In order to have a late time
behavior consistent with Lloyd's bound one is forced to  have a cut off behind the horizon 
whose value is  fixed by the boundary cut off.  Using this result we compute holographic 
complexity for two dimensional AdS solutions where we get expected  late times linear growth. 
 It is in contrast with the naively computation which is done without assuming the cut off where the
 complexity approaches a constant at the late time.

\end{abstract}

\end{titlepage}

\section{Introduction}

 According to the  ``complexity=action'' proposal  (CA) the quantum computational complexity of a 
 holographic state is given by the on-shell action evaluated on a bulk region known as the 
 `Wheeler-De Witt' (WDW) patch \cite{Brown:2015bva, Brown:2015lvg}
 \be
{\cal C}(\Sigma)=\frac{I_{\rm WDW}}{\pi \hbar}.
\ee
Here the WDW patch is defined as the domain of dependence of any Cauchy surface in the bulk 
whose intersection with the asymptotic boundary is the time slice $\Sigma$. 
 
 An  interesting feature of the complexity is that it grows linearly with time at the late time with
 slope given by Lloyd's bound \cite{Lloyd:2000}  that is twice of the energy of the state.
 Holographic complexity for two-sided black holes has been {calculated} in \cite{Carmi:2017jqz} where
it was shown that although at the late time the growth rate  approaches a constant value that is 
twice of the mass of the black hole, the constant is approached from above, violating the Lloyd's 
bound \cite{Lloyd:2000}.

Another recent interesting development in the literature of theoretical higher energy is to study 
a conformal theory deformed by an irrelevant operator such as the one which is quadratic 
in the stress energy tensor known as $T\overline{T}$ deformation.  Although typically deforming 
a conformal field theory by an irrelevant operator would remove UV fixed point and 
makes it non-local at high energies, it was shown that for the mentioned deformation 
the resultant theory is still exactly solvable\cite{{Smirnov:2016lqw},{Cavaglia:2016oda}}.

To be concrete let us consider a two dimensional conformal field theory deformed by the 
corresponding operator as follows
\be
I_{\rm QFT}=I_{\rm CFT}+\mu\int d^2x T\overline{T}.
\ee
There are some interesting features of the resultant quantum field theory. First of all it is 
UV complete. Moreover the spectrum of the deformed theory can be  determined 
 non-perturbatively and rather in a compact form.  More precisely for a conformal field theory
 on a cylinder with the circumference $L$ the energy level $E_n(\mu,L)$  for a
 state denoted by conformal dimensions $(\Delta_n,\bar{\Delta}_n)$ is 
 given by \cite{{Smirnov:2016lqw},{Cavaglia:2016oda}}  
  \be 
 E_n(\mu,L)=\frac{2L}{\mu}\left(1-\sqrt{1-\frac{2\pi \mu}{L^2}\left(M_n+\frac{2\pi \mu}{L^2}J_n^2
 \right)}
\hspace{2mm} \right),
 \ee
 where $M_n=\Delta_n+\bar{\Delta}_n-\frac{c}{12}$, and $J_n=\Delta_n-\bar{\Delta}_n$.

In the context of AdS/CFT correspondence it was proposed that the above deformation 
has a holographic dual. The corresponding dual gravitational  theory may be described 
 by  an AdS$_3$ 
metric  with a finite radial cut off\cite{McGough:2016lol}. 
The radial cut off $r_c$ is given in terms of the deformed 
parameter $\mu$,  by $r_c^2=\frac{16\pi G}{\mu}$.

Using AdS/CFT correspondence the generalization of $T\overline{T}$ deformation
to higher dimensional conformal field theories has also been studied in 
\cite{{Taylor:2018xcy},{Hartman:2018tkw}}. Following \cite{McGough:2016lol}
one would also expect that  a $d+2$ dimensional AdS black brane solution with a radial cut off 
could provide a holographic dual for a $d+1$ dimensional  $T\overline{T}$ deformed conformal 
field theory. Given the corresponding geometry by 
\be\label{metric}
ds^2=\frac{\ell^2}{r^{2}}
\left(-{f(r)}dt^2
+\frac{dr^2}{f(r)}+\sum_{i=1}^d d\vec{x}^2\right),\;\;\;\;\;\;\;\;f(r)=1-\left(\frac{r}{r_h}\right)^{d+1},
\ee
where  $r_h$ and  $\ell$ are radius of horizon  the AdS radius, respectively, the spectrum of 
energy  of the deformed theory is \cite{{Taylor:2018xcy},{Hartman:2018tkw}}
\be\label{EE}
E=\dfrac{V_d \ell^d d}{8\pi G}\frac{1}{r^{d+1}_{c}}\bigg(1-\sqrt{1-
\frac{r_c^{d+1}}{r_h^{d+1}}}\hspace{1mm}\bigg),
\ee
with  $V_d$ being the volume of $d$-dimensional internal space of the metric parametrized by 
$x_i, \,i=1,\cdots d$.

Motivated by  $T{\overline T}$ deformation  and its holographic dual in the present paper we 
would like to compute the  complexity growth of a black brane at a finite cut off using 
CA proposal. We observe  that requiring  to reach the  Lloyd's bound at the late time 
enforces us to have  a cut off behind the horizon whose value is fixed by boundary cut off. More 
precisely for  black brane solutions denoting  cut off radius  inside the horizon by $r_0$ 
one finds (at leading order)
\be\label{Cutoff}
r_0r_c^2= 2^{\frac{4}{d+1}}{r_h^3}.
\ee
To explore the significance of our result we will then study holographic complexity 
for AdS$_2$ vacuum solutions  of certain two dimensional Maxwell-Dilaton gravities. 
One observes that if we naively  compute the complexity without taking into account the
behind horizon cut off the  rate of growth  vanishes at  the late time. On the other had 
if one assumes that the UV cut off would set a cut off behind the horizon given by 
\eqref{Cutoff} the complexity exhibits late time linear growth, as expected.

The paper is organized as follows. In the next section we will compute holographic complexity
for back brane solutions in the present of a cut off where we show how the inside cut off 
would emerge. In section three we will study complexity for AdS$_2$ taking into account the 
enforced behind  the horizon cut off. The last section is devoted to conclusions.

\section{CA complexity for cut off geometries}

In this section we would like to compute holographic complexity for a black brane solution with 
a radial cut off. To do so, following CA proposal  we will need to compute on shell action on 
the WDW patch associated with 
a boundary state given at $\tau=t_L+t_R$. Here $t_L (t_R)$, is  time coordinate of left (right)
boundary on the eternal black brane (see figure 1). Of course since we are interested in the 
late time behavior of the complexity it is sufficient  to compute on shell action over 
 the intersection of the WDW patch with the  
future interior shown by dark blue color in the  figure 1\cite{Alishahiha:2018lfv}.
\begin{figure}
\begin{center}
\includegraphics[width=0.4\linewidth]{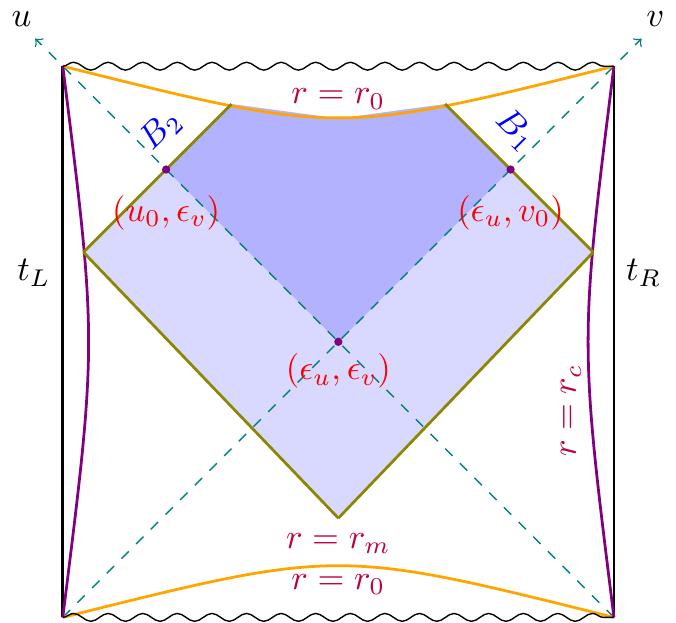}
\caption{The intersection of WDW patch with the future interior of an eternal AdS black brane. 
The theory is defined at a radial finite cut off $r_c$ that fixes a cut off behind the horizon 
denoted by $r_0$. }
\end{center}
\end{figure}\label{Fig1}

To proceed we note that the action consists of several parts  including bulk, boundary and joint 
points as follows \cite{{Parattu:2015gga},{Parattu:2016trq},{Lehner:2016vdi}}
\bea\label{ACT0}
I^{(0)}&=&\frac{1}{16\pi G_N}\int d^{d+2}x\sqrt{-g}(R-2\Lambda)+\frac{1}{8\pi G_N}
\int_{\Sigma^{d+1}_t} K_t\; d\Sigma_t\cr &&\cr
&& \pm\frac{1}{8\pi G_N} \int_{\Sigma^{d+1}_s} K_s\; d\Sigma_s\pm \frac{1}{8\pi G_N} 
\int_{\Sigma^{d+1}_n} K_n\; dS 
d\lambda \pm\frac{1}{8\pi G_N} \int_{J^d} a\; dS\,.
\eea
Here the timelike, spacelike, and null boundaries and also joint points are denoted by $
\Sigma_t^{d+1}, \Sigma_s^{d+1}, \Sigma_n^{d+1}$ and $J^d$, respectively. The extrinsic 
curvature of the corresponding boundaries are given by $K_t, K_s$ and $K_n$. The function $a$ 
at the intersection of the boundaries is given by the logarithm of the inner product of the 
corresponding normal vectors and $\lambda$ is the null coordinate defined on 
the null segments. The sign of different terms depends on the relative position 
of the boundaries and the bulk region of interest (see \cite{Lehner:2016vdi} for more details).

The null boundaries $B_1$ and $B_2$ of the  future interior are  
\be
B_1:\,\,t=t_R+r^*(r_c)-r^*(r),\;\;\;\;\;\;\;\;\;\;B_2:\,\,t=-t_L-r^*(r_c)+r^*(r),
\ee
where $r^*(r)$ is the tortoise coordinate. The null vectors associated with these null boundaries
 are also given by
 \be
 k_1=\alpha \left(\partial_t+\frac{1}{f(r)}\partial_r\right),\;\;\;\;\;\;\;\;\;\;\; k_2=\beta\left(\partial_t-
 \frac{1}{f(r)}\partial_r\right), 
 \ee
 Here $\alpha$ and $\beta$ are two free constant parameters appearing due to the 
 ambiguity of the normalization of null vectors. 
 
 Using this notation  the bulk part of the on shell action is 
\bea
I^{\rm bulk}_{\rm FI}&=&-\frac{V_{d}\ell^d}{4\pi G_N}(d+1)\int_{r_h}^{r_{0}} \frac{dr}{r^{d+2}}
\left(\frac{\tau}{2}+r^*(r_c)-r^*(r)\right)\cr &&\cr
&=&-\frac{V_{d}\ell^d}{8\pi G_N}\left(\frac{2}{d\,r_h^d}-\frac{2}{d\,r_0^d}\right)
-\frac{V_{d}\ell^d}{8\pi G_N}\left(\frac{1}{r_h^{d+1}}
-\frac{1}{r_0^{d+1}}
\right)(\tau+\tau_c).
\eea
where $\tau_c=2(r^*(r_c)-r^*(r_0))$. Here to find the last expression  we have performed an 
integration by parts.

There are five boundaries four of which are null that have zero contribution if one uses the 
Affine parameter  to parametrize the null directions. Therefore we are left with a 
space like boundary at  future singularity whose contribution is given by
\be\label{surf1}
I^{\rm surf_1}_{\rm FI}=-\frac{1}{8\pi G_N}\int d^dx\, \int_{-t_L-r^*(r_c)+r^*(r)}^{
t_R+r^*(r_c)-r^*(r)} dt\, \sqrt{h}K_s 
\Big|_{r=r_{0}},
\ee
where $K_s$ is the the trace of extrinsic curvature of the boundary at $r=r_{0}$ and $h$ 
is the determinant of the induced metric on it. To compute this term it is 
useful to note that for a constant $r$ surface using the metric \eqref{metric} one has
\be
\sqrt{h} K=-\sqrt{g^{rr}}\partial_r\sqrt{h}=-\frac{1}{2}\frac{\ell^d}{r^{d}}\left(\partial_r f(r)-
\frac{2(d+1)}{r}f(r)\right),
\ee
therefore the boundary term \eqref{surf1} reads 
\bea\label{SURF}
I^{\rm surf_1}_{\rm  FI}=\frac{V_d\ell^d}{8\pi G_N }(d+1)\left(\frac{1}{2r_h^{d+1}}
-\frac{1}{r_0^{d+1}}\right)
(\tau+\tau_c)\,.
\eea
Note that there is also another boundary term to be evaluated at the surface 
cut off behind the horizon that is  given by
\be\label{surf2}
I^{\rm surf_2}_{\rm FI}=\frac{1}{8\pi G_N}\int d^dx\, \int_{-t_L-r^*(r_c)+r^*(r)}^{
t_R+r^*(r_c)-r^*(r)} dt\, \sqrt{|h|}\,\,\,\frac{d}{\ell}\,
\Big|_{r=r_{0}}=\frac{V_d\ell^d}{8\pi G_N}\frac{d}{r_0^{d+1}}\,\sqrt{\frac{r_0^{d+1}}
{r_h^{d+1}}-1}\,\,(\tau+\tau_c)\,.
\ee

There are also  five joint points,  two points at   $r_{0}$ and three  at the horizon $r=r_h$.
Of course those at the horizon are not at the same point, though the coordinate system
$r$ cannot make any distinction between them. To label these points it is  convenient to use the
following  coordinate system \cite{Agon:2018zso}, 
\bea
u=-e^{-\frac{1}{2}f'(r_h)(r^*(r)-t)},\;\;\;\;\;\;\;\;\;\;\;\; v=-e^{-\frac{1}{2}f'(r_h)(r^*(r)+t)}\,,
\eea
by which the  points are located at  $(\epsilon_u,v_0)$, $(u_0,\epsilon_v)$ 
and $(\epsilon_u,\epsilon_v)$ as depicted in figure 1. Here in order to  regularize 
quantities like $r^*(r)$ at $r=r_h$ we have put the horizon at $v=\epsilon_v$ and $u=\epsilon_u$
for small $\epsilon_v$ and $\epsilon_u$.
In what follows the radial coordinate associated with these three
points are also labeled  by $r_{v_0}, r_{u_0}$ and $r_\epsilon$, respectively. Using this notation 
the contribution of joint points is \cite{Alishahiha:2018lfv}
\bea\label{joint-3} 
I^{\rm joint}_{\rm FI}&=&\frac{V_d \ell^d}{8\pi G_{N}}
\left(\frac{\log\frac{\alpha\beta r_0^2}{\ell^2|f(r_0)|}}{r_0^{d}}
+\frac{\log\frac{\alpha\beta r_\epsilon^2}{\ell^2|f(r_\epsilon)|}}{r_\epsilon^{d}}
-\frac{\log\frac{\alpha\beta r_{u_{0}}^2}{\ell^2|f(r_{u_{0}})|}}{r_{u_0}^{d}}
-\frac{\log\frac{\alpha\beta r_{v_{0}}^2}{\ell^2|f(r_{v_{0}})|}}{r_{v_0}^{d}}
\right)\\&&\cr
&=&-\frac{V_d \ell^d}{8\pi G_{N}}
\left(\frac{\log{|f(r_{\epsilon})|}-\log{|f(r_{u_{0}})|}-\log{|f(r_{v_{0}})|}}{r_{h}^{d}}
+\frac{\log\frac{\alpha\beta r_{h}^2}{\ell^2}}{r_{h}^{d}}+\frac{\log\frac{\alpha\beta r_0^2}{\ell^2|
f(r_0)|}}{r_0^{d}}\right)\,.\nonumber
\eea
Here we have used the fact that $\{r_{u_m},r_{v_m},r_{\epsilon}\}\approx r_h$. On the 
other hand by making use of the fact that \cite{Agon:2018zso}
\be\label{Hlim}
\log |f(r_{u,v})|=\log|uv|+ c_0+{\cal O}(uv)\;\;\;\;\;\;\;\;\;\;\;{\rm for}\;\;uv\rightarrow 0,
\ee
 one arrives at 
 \be
I^{\rm joint}_{\rm FI}=\frac{V_d \ell^d}{8\pi G_{N}}
\left(\frac{\log |u_{0} v_{0}|+c_0}{r_{h}^{d}}-\frac{\log |f(r_0)|}{r_0^{d}}\right)
-\frac{V_d \ell^d}{8\pi G_{N}}\left(\frac{\log\frac{\alpha\beta r_{h}^2}{\ell^2}}{r_{h}^{d}}
-\frac{\log\frac{\alpha\beta r_0^2}{\ell^2}}{r_0^{d}}\right).
\ee

The only remaining part of the action to be considered is a term needed to remove the ambiguity
associated with the normalization of null vectors \cite{{Lehner:2016vdi},{Reynolds:2016rvl},
{Alishahiha:2018tep}}
\be
I^{\rm amb}=\frac{1}{8\pi G_N}\int d\lambda d^dx \sqrt{\gamma}\Theta\log\frac{|\tilde{\ell}\Theta|}{d},
\ee
where $\tilde{\ell}$ is an undetermined  length scale and  
$\gamma$ is the determinant of the induced 
metric on the joint point where two null segments intersect, and 

\be
\Theta=\frac{1}{\sqrt{\gamma}}\frac{\partial\sqrt{\gamma}}{\partial\lambda}\,.
\ee 
In the the present  case  the contribution of this term is 
(for more details see \cite{Alishahiha:2018lfv})
\be
I^{\rm amb}_{\rm FI}=\frac{V_d\ell^d}{8\pi G_N}\left(\frac{\log\frac{\alpha\beta \tilde{\ell}^2r_h^2}
{\ell^4}}{r_h^{d}}-\frac{\log\frac{\alpha\beta \tilde{\ell}^2r_0^2}
{\ell^4}}{r_0^{d}}\right)+\frac{V_d\ell^d}{8\pi G_N}\left(\frac{2}{d\,r_h^{d}}-
\frac{2}{d\,r_0^{d}}\right).
\ee
Taking all parts contributing to the on shell action into account one arrives at
\bea
I_{\rm FI}\!\!&\!\!=\!\!&\!\!\frac{V_d \ell^d }{8\pi G_{N}}\Bigg[\left(
\frac{d}{r_h^{d+1}}
-\frac{d}{r_0^{d+1}}+\frac{d}{r_0^{d+1}}\,\sqrt{\frac{r_0^{d+1}}{r_h^{d+1}}-1}
\right)(\tau+\tau_c)+\frac{(d+1)r^*(r_0)+c_0r_h}{r_{h}^{d+1}}-\frac{\log |f(r_0)|}{r_0^{d}}
\cr &&\cr &&\;\;\;\;\;\;\;\;\;\;\;\;\;+\left(\frac{1}{r_h^{d}}-\frac{1}{r_0^{d}}\right)
\log\frac{ \tilde{\ell}^2}{\ell^2}
\Bigg],
\eea
leading to the following rate of growth
\be
\frac{dI_{\rm FI}}{d\tau}
=\frac{V_d \ell^d d}{8\pi G_{N}}\left(
\frac{1}{r_h^{d+1}}-\frac{1}{r_0^{d+1}}+\frac{1}{r_0^{d+1}}\,\sqrt{\frac{r_0^{d+1}}{r_h^{d+1}}-1}\right),
\ee
that is indeed the late time expression for the holographic  complexity of the corresponding 
black brane solution \cite{Alishahiha:2018lfv}. Now the aim is to compare the above result with 
Lloyd's bound that in our case should be read from the energy spectrum \eqref{EE}
that can be recast into the following inspiring form
\be
E=\dfrac{V_d \ell^d d}{16\pi G}\frac{1}{r^{d+1}_{h}}+\dfrac{V_d \ell^d d}{16\pi G}
\frac{1}{r^{d+1}_{c}}\bigg(1-\sqrt{1-
\frac{r_c^{d+1}}{r_h^{d+1}}}\bigg)^2\,.
\ee
Therefore if one assumes that at the late time the growth rate of complexity saturates the Lloyd's 
bound, $2E$,  one may conclude that 
\be
\frac{1}{r_0^{d+1}}\left(\sqrt{\frac{r_0^{d+1}}{r_h^{d+1}}-1}-1\right)=
\frac{1}{r^{d+1}_{c}}\bigg(\sqrt{1-
\frac{r_c^{d+1}}{r_h^{d+1}}}-1\bigg)^2,
\ee
which at leading order reduces to
\be
r_0r_c^{2}=
2^{\frac{4}{d+1}}r_h^{3},
\ee
This  means that the cut off at the singularity is fixed by the UV cut off at  the boundary.  
In other words this leads to a conclusion that as soon as we fixed the UV cut off we are not allowed 
to consider another independent cut off inside the horizon (let say near the singularity) and 
the UV cut off will automatically regularize the modes inside the horizon. 
This is, indeed,  the main result of the present paper. 

To explore the importance of the above conclusion in what follows we will study
holographic complexity for AdS$_2$ vacuum solutions  of certain two dimensional gravities.


\section{Complexity for AdS$_2$ geometry}

In this section we shall study holographic complexity for certain  two dimensional 
Maxwell-Dilaton gravities that admit AdS vacuum solutions. The first model we will consider 
has the following action\footnote{This is indeed one of the 
simplest example of two dimensional gravity having non-trivial vacuum. One could, as well,
consider rather  more complicated actions (see {\it e.g.} \cite{{Alishahiha:2008rt},{Guralnik:2003we},{Grumiller:2003ad}}).   }
\be\label{actions} 
I=\frac{1}{8G}\int d^2x\sqrt{-g} \left(e^{\phi}\left(R+\frac{2}{\ell^2}\right)-F^2\right).
\ee
Using the entropy function formalism \cite{Sen:2005wa} one can show that  the above action 
 admits constant dilaton AdS$_2$ vacuum solution as follows \cite{Alishahiha:2008tv}
 \be
ds^2={\ell^2}\left(-(r^2-r_h^2) dt^2+\frac{dr^2}{r^2-r_h^2}\right),\;\;\;\;\;e^{\phi}=
4G^2Q^2\ell^2,\;\;\;\;\;
F_{rt}=2GQ\ell^2,
\ee
whose entropy is  
\be
S_{\rm BH}=2\pi GQ^2\ell^2,
\label{BHE}
\ee 
that it is independent of $r_h$.  Let us compute holographic complexity for a state
given at $\tau=t_L+t_R$. The corresponding WDW patch is depicted in the figure 2.
\begin{figure}\label{Fig:AdS2}
\begin{center}
\includegraphics[width=0.24\linewidth]{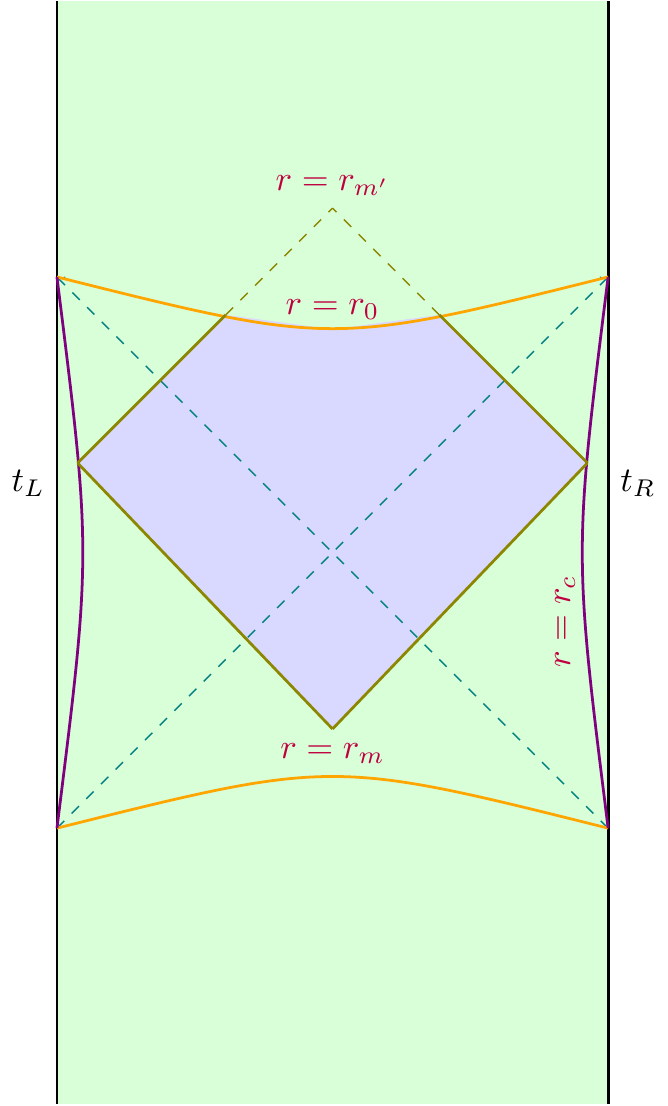}
\caption{Penrose diagram of AdS$_2$ geometry. The green part is covered by AdS global
coordinates, while the Rindler coordinates cover a portion shown in the figure. The 
actual WDW patch is shown by blue color.}
\end{center}
\end{figure}

One may naively compute on shell action in the WDW patch shown in the figure 2 
with two joint points denoted by $r_m$ and $r_{m'}$ ( the later point is drown by dashed lines).  
Positions of the corresponding points are obtained from the following equations
\be
\tau=-2(r^*(r_c)-r^*(r_{m'}))=2(r^*(r_c)-r^*(r_{m})),
\ee
where $r_c$ is a UV cut off.

Following CA proposal the idea is to evaluate on shell action on the corresponding WDW with  a 
UV cut off but no, a priori,  restriction on modes behind the horizon. This means that there is 
no cut off behind the horizon and  both corners denoted by $m$ and $m'$ should be 
taken into account. With this assumption the bulk part of the on shell action reads
\bea
I^{\rm bulk}&=&G Q^2\ell^2\bigg(\int_{r_{m'}}^{r_h} dr \left(\tau+2(r^*(r_c)-r^*(r))\right)
+2\int^{r_{c}}_{r_h} dr \, 2\left(r^*(r_c)-r^*(r)\right)\cr &&\cr &&
\;\;\;\;\;\;\;\;\;\;\;\;\;\;\;\;\;+\int_{r_{m}}^{r_h} dr \left(-\tau+2(r^*(r_c)-r^*(r))\right)\bigg)
\cr &&\cr
&=&2G Q^2\ell^2\bigg( (r_{m}-{r_{m'}}  )\frac{\tau}{2}+\int_{r_{m'}}^{r_c} dr(r^*(r_c)-r^*(r))
+\int_{r_{m}}^{r_c} dr \, \left(r^*(r_c)-r^*(r)\right)\bigg)\,.
\eea
By making use of an integration by parts one finds
\be
I^{\rm bulk}=GQ^2\ell^2\bigg(2\log | f(r_c)|-\log | f(r_m)|-\log | f(r_{m'})|\bigg)\,,
\ee
where $f(r)=r^2-r_h^2$\,.
 
On the other hand using Affine parameter to parametrize the null direction one gets zero
contribution from null boundaries.  Therefore the only part one needs to further consider 
 is the contribution of join points. Denoting the null vectors by
\be
k_1=\alpha\left(\partial_t-\frac{1}{f(r)}\partial_r\right),\;\;\;\;\;\;\;\;k_2
=\beta\left(\partial_t+\frac{1}{f(r)}\partial_r\right),
\ee
one gets
\bea
I^{\rm joint}&=&\frac{e^\phi}{4G}\left(\log \bigg|\frac{\alpha\beta}{\ell^2 f(r_m)}\bigg|+
\log \bigg|\frac{\alpha\beta}{\ell^2 f(r_{m'})}\bigg|-2\log \bigg|\frac{\alpha\beta}{\ell^2 f(r_c)}\bigg|\right)\cr &&\cr
&=&GQ^2\ell^2\bigg(2\log | f(r_c)|-\log | f(r_m)|-\log | f(r_{m'})|\bigg).
\eea
Interestingly enough the free parameters $\alpha$ and $\beta$ drop from the final result which 
means that there is no ambiguity associated with the normalization of null vectors. 
Therefore we do not
need any further counter terms, except possibly the one that could cancel the most divergent 
term of the on shell action, $\log f(r_c)$. Of course since we are interested in the time derivative 
of the action this term does not play any role.

Taking all terms contributing to the on shell action  one arrives at
\be
I=I^{\rm bulk}+I^{\rm joint}=2GQ^2\ell^2\bigg(2\log | f(r_c)|-\log | f(r_m)|-\log | f(r_{m'})|\bigg),
\ee
whose time derivative is 
\be
\frac{dI}{d\tau}=2GQ^2\ell^2 (r_m-r_{m'}).
\ee
It is then notable that  at the late time when $\{r_m,r_{m'}\}\rightarrow r_h$ the rate of growth 
vanishes, leading to a constant late time complexity that is counter intuitive. Indeed we would 
expect  to get linear growth at the late time.
 
Of course in light of our result in the previous section this  conclusion is, indeed,  misleading.
In fact,  as we have already demonstrated in the previous section, setting a UV cut off at the 
boundary  would enforce us to have a cut off inside the horizon that prevents us to have
access to all regions on WDW located behind the horizon.

In other words,  as soon as we set the UV  cut off, $r_c$, at the boundary we will also have 
to consider  a cut off behind the horizon given by  $r_0 \sim \frac{{r_h^3}}{r_c^2}$ at leading order. Actually 
having this cut off will remove the joint point $r_{m'}$ from the 
WDW patch and instead we would have a space like boundary at $r=r_0$. Therefore one should 
redo our computations for on shell action for a new WDW patch  that has no  
joint point $m'$, as shown with blue color in the figure 2.

To proceed let us again start with the bulk action. In this case one gets
 \bea
I_{\rm bulk}&=&G Q^2\ell^2\bigg(\int_{r_{0}}^{r_h} dr \left(\tau+2(r^*(r_c)-r^*(r))\right)
+2\int^{r_{c}}_{r_h} dr \, 2\left(r^*(r_c)-r^*(r)\right)\cr &&\cr &&
\;\;\;\;\;\;\;\;\;\;\;\;\;\;\;\;\;+\int_{r_{m}}^{r_h} dr \left(-\tau+2(r^*(r_c)-r^*(r))\right)\bigg),
\eea 
that can be recast to the following form after making use of an integration by parts
\be
I_{\rm bulk}=GQ^2\ell^2\bigg(2\log | f(r_c)|-\log | f(r_m)|-\log | f(r_{0})|-r_0\left(\tau+2
(r^*(r_c)-r^*(r_0))\right)\bigg)\,.
\ee
The boundary contributions associated with null boundaries are still zero when Affine
parametrization is used. Of course in the present case we have a apace like boundary whose 
contribution is
\be
I_{\rm surf}=-\frac{1}{4G}\int dt e^{\phi}\sqrt{-h}\left(K_s-\frac{1}{\ell}\right)\bigg|_{r_0}=
GQ^2\ell^2 (r_0+r_h)
\left(\tau+2
(r^*(r_c)-r^*(r_0))\right)\,.
\ee
As for joint points we have 
\bea
I_{\rm joint}&=&\frac{e^\phi}{4G}\left(\log \bigg|\frac{\alpha\beta}{\ell^2 f(r_m)}\bigg|+
\log \bigg|\frac{\alpha}{\ell \sqrt{f(r_{0})}}\bigg|+\log \bigg|\frac{\beta}{\ell \sqrt{f(r_{0})}}\bigg|
-2\log \bigg|\frac{\alpha\beta}{\ell^2 f(r_c)}\bigg|\right)\cr &&\cr
&=&GQ^2\ell^2\bigg(2\log | f(r_c)|-\log | f(r_m)|-\log | f(r_0)|\bigg).
\eea
Now putting all terms together one arrives at
\be
I=2GQ^2\ell^2\bigg(2\log | f(r_c)|-\log | f(r_m)|-\log | f(r_{0})|\bigg)+GQ^2\ell^2 r_h
\left(\tau+2
(r^*(r_c)-r^*(r_0))\right)\,.
\ee
It is the easy to show 
\be
\frac{dI}{dt}=GQ^2\ell^2(r_h+2 r_m),
\ee
which approaches a constant at the late time
\be
\frac{dI}{dt}=3GQ^2\ell^2 r_h=3 (2\pi G Q^2\ell^2)(\frac{r_h}{2\pi})=3S_{\rm BH} T\,.
\ee
Here $T$ is the Hawking temperature associated with the geometry. This is in 
agreement with what is expected; namely one has late time linear growth with slop
given by  entropy times temperature. Of course the actual numerical factor does not look
universal. 

To further explore the above picture better it is also constructive to consider another 
two dimensional model admitting AdS$_2$ vacuum solutions as follows
\be
S_2=\frac{1}{8G}\int d^2x\sqrt{-g}\; e^{\phi}\left(R+\frac{2}{\ell^2}
-\frac{\ell^2}{4}e^{2\phi}F^2\right).
\label{S2}
\ee
Using the entropy function formalism  \cite{Sen:2005wa} one can show that  the above action 
 admits the AdS$_2$ vacuum solution as follows \cite{Alishahiha:2008tv}\footnote{See
 \cite{Cvetic:2016eiv} for non-constant dilation solution of the model.}
 \be
ds^2=\frac{\ell^2}{4}\left(-(r^2-r_h^2)dt^2+\frac{dr^2}{r^2-r_h^2}\right),\;\;\;\;\;\;e^{\phi}=\sqrt{4GQ\ell^2},\;\;\;\;\;\;
F_{tr}=\sqrt{\frac{1}{16GQ\ell^2}}
\ee
with the entropy, 
\be
S_{\rm BH}=2\pi\sqrt{\frac{Q\ell^2}{4G}}.
\label{ccc}
\ee 
Now the aim is to compute the holographic complexity for this model. Of course the procedure 
is the same as that we considered  in the previous case and the only difference is the numerical
factors. More precisely for the bulk term one finds
\be
I_{\rm bulk}=-\frac{\ell}{4}\sqrt{\frac{Q}{G}}\bigg(2\log | f(r_c)|-\log | f(r_m)|-
\log | f(r_{0})|-r_0\left(\tau+2
(r^*(r_c)-r^*(r_0))\right)\bigg)\,.
\ee
As for joint points one gets
\bea
I_{\rm joint}=\frac{\ell}{2}\sqrt{\frac{Q}{G}}\bigg(2\log | f(r_c)|-\log | f(r_m)|-\log | f(r_0)|\bigg),
\eea
while for the surface term one has
\be
I_{\rm surf}=\frac{\ell}{2}\sqrt{\frac{Q}{G}}\,(r_0+r_h)
\left(\tau+2
(r^*(r_c)-r^*(r_0))\right)\,.
\ee
 Therefore the total action is found
\be
I=\frac{\ell}{4}\sqrt{\frac{Q}{G}}\bigg(2\log | f(r_c)|-\log | f(r_m)|-\log | f(r_0)|
\bigg)+\frac{\ell}{2}\sqrt{\frac{Q}{G}}\,r_h
\left(\tau+2(r^*(r_c)-r^*(r_0))\right),
\ee
resulting to the following rate of growth for the on shell action 
\be
\frac{dI}{dt}=\frac{\ell}{4}\sqrt{\frac{Q}{G}}\, (r_m+2 r_h),
\ee
which  approaches a constant at late time
\be
\frac{dI}{dt}=3\frac{\ell}{4}\sqrt{\frac{Q}{G}}\,r_h=\frac{3}{2} S_{\rm BH} T\,.
\ee
Note that the same as  previous one,  had not we considered the inside surface cut off, 
the complexity growth would have been zero at the late time.


\section{Conclusions} 

In this paper we have studied holographic complexity for an AdS black brane geometry with 
a radial cut off using CA proposal. Within  this explicit example we have found that
as soon as one sets a UV cut off at the boundary the model enforces us to have a cut off 
behind the horizon whose value is fixed by the UV cut off. Indeed  in the present case 
one has
\be
\frac{1}{r_0^{d+1}}\left(\sqrt{\frac{r_0^{d+1}}{r_h^{d+1}}-1}-1\right)=
\frac{1}{r^{d+1}_{c}}\bigg(\sqrt{1-
\frac{r_c^{d+1}}{r_h^{d+1}}}-1\bigg)^2\,.
\ee
It is worth mentioning that in order to get a consistent result fulfilling the Lloyd's bound 
it was crucial to consider the contribution of certain counter term on the cut off surface 
behind the horizon.

In this paper we have only considered uncharged black hole with flat boundary. It would be 
interesting to find an expression for  behind the horizon cut off in terms of the UV cut off 
for a general charged black hole. In general the cut off $r_0$ is a function of UV cut off;
$r_0=r_0(r_h,r_c)$, though it might not  have such a simple expression as  above. 
 Actually this relation should be intuitively understood from the fact  
that the energy is a charge defined at the boundary while the late time behavior of 
complexity is evaluated from the action behind the horizon.

If our result works for a generic black hole, it means that  near singularity  modes 
may be regularized through a UV cut off. It is, however,  important to note that our conclusion 
will not affect the results people have found so far in the literature, though it might shed light on 
some new problem such as how to deal with Riemann tensor squared.

Actually in order to explore the importance of our result we have studied holographic complexity 
for AdS$_2$ vacuum solutions in certain two different Maxwell-Dilaton gravities. We have found 
that the complexity is finite at late times if one does not consider the cut off enforced by 
the UV cut off, that seems  counter intuitive. Indeed  one would expect that the complexity 
exhibits  linear growth at the late time. On the other hand if one considers behind the horizon 
cut off fixed by the UV cut off,  indeed one  gets the corresponding linear growth.

Two dimensional AdS solutions we have considered were supported by a constant Dilaton,
though it would be interesting to consider the case where the Dilaton is not constant. 
This might be more interesting as it could provide a holographic dual for SYK model
\cite{{Sachdev:1992fk},{Kitaev}} (see for example \cite{{Jensen:2016pah},{Maldacena:2016upp}
,{Engelsoy:2016xyb}}).

{\bf Note added:} While we were in the final stage of our work, the paper \cite{Brown:2018bms}
appeared in the arXiv where the complexity of two dimensional gravity has also been studied. 
In this paper the authors resolved the undesired late time behavior  by adding a new charge 
to the model. This in fact could be naturally accommodated if one considers the model as a 
dimensionally  reduced four dimensional RN black hole.

\subsection*{Acknowledgements}
The authors would like to kindly thank K. Babaei, A. Faraji Astaneh, G. Jafari, M. R. Mohammadi Mozaffar,
 F. Omidi, M. R. Tanhayi and  M.H. Vahidinia for useful comments  and 
discussions  on related topics.  We would also like to thank Leonard Susskind for 
a correspondence.

\end{document}